# Eulerian and Newtonian dynamics of quantum particles

S.A. Rashkovskiy

*Institute for Problems in Mechanics, Russian Academy of Sciences, Vernadskogo Ave., 101/1 Moscow, 119526, Russia, Tel. +7 495 5504647, E-mail: rash@ipmnet.ru*

We derive the classical equations of hydrodynamics (the Euler and continuity equations), from which the Schrödinger equation follows as a limit case. It is shown that the statistical ensemble corresponding to a quantum system and described by the Schrödinger equation can be considered an inviscid gas that obeys the ideal gas law with a quickly oscillating sign-alternating temperature. This statistical ensemble performs the complex movements consisting of smooth average movement and fast oscillations. It is shown that the average movements of the statistical ensemble are described by the Schrödinger equation. A model of quantum motion within the limits of classical mechanics that corresponds to the hydrodynamic system considered is suggested.

## I. INTRODUCTION

The methods used to describe the motion of particles in classical and quantum mechanics are fundamentally different. It is well known that there are several alternative formulations of classical mechanics [1,2], from which, for our purposes, we highlight only two: the Newtonian formulation and the Hamilton-Jacobi theory.

In the Newtonian formulation of classical mechanics, the motion of a particle is described by Newton's second law (or its alternative records in the form of, e.g., Hamilton or Lagrange equations):

$$m\ddot{\mathbf{r}} = -\nabla U \qquad (1)$$

which allows the calculation of the particle trajectory $\mathbf{r}(t, \mathbf{r}_0, \mathbf{v}_0)$ at the prescribed initial conditions of

$$\mathbf{r}(t=0) = \mathbf{r}_0, \quad \dot{\mathbf{r}}(t=0) = \mathbf{v}_0 \qquad (2)$$

and, thus, at each instant, specifies at which point of space the given particle is located.

In the Hamilton-Jacobi theory, the motion of an ensemble of identical non-interacting particles, which we call the Hamilton-Jacobi ensemble, is considered rather than the motion of a single particle. This ensemble is characterized by a density $\rho(\mathbf{r},t)$, which satisfies the continuity equation:

$$\frac{\partial \rho}{\partial t} + \mathrm{div}(\rho\mathbf{v}) = 0 \qquad (3)$$

where

$$\mathbf{v}(\mathbf{r},t) = \frac{1}{m}\nabla S \quad (4)$$

is the velocity field of the ensemble, and the function $S(\mathbf{r},t)$, which has the sense of action, satisfies the classical Hamilton-Jacobi equation:

$$\frac{\partial S}{\partial t} + \frac{1}{2m}|\nabla S|^2 + U = 0 \quad (5)$$

The trajectory of an individual particle in the Hamilton-Jacobi theory can be found either using Jacobi's theorem [1,2] with the complete integral of Eq. (5) or by solving the system of ordinary differential equations:

$$\dot{\mathbf{r}} = \frac{1}{m}\nabla S(\mathbf{r},t) \quad (6)$$

using a solution of the Hamilton-Jacobi equation (5).

The interconnection of both formulations of classical mechanics is well known and obvious: using the Hamilton-Jacobi equation (3) as a source and separating the potential energy, we can construct Newton's equations (1), and vice versa. Using a solution of Newton's equations (1), we can construct a complete integral of the Hamilton-Jacobi equation (3) [1,2]. From a mathematical perspective, equation (1) describes the characteristics of the Hamilton-Jacobi equation (5).

In quantum mechanics, the motion of a particle is described by a wave function $\psi(\mathbf{r},t)$, which is the solution of the Schrödinger equation. The classical notion of the trajectory of a single particle in quantum mechanics is meaningless, and it is only possible to discuss the probability of finding the particle at different points in space but not how the particle came to a particular point. According to Born's probabilistic interpretation [3,4], the probability density of finding the particle at a given point is proportional to $|\psi(\mathbf{r},t)|^2$. On this basis, one can argue that, in quantum mechanics similarly to the classical Hamilton-Jacobi theory, we do not consider a single particle but rather an ensemble of identical non-interacting particles, which can be called the Schrödinger ensemble.

Numerous attempts to construct *a quantum mechanics of individual particles* using classical concepts, including "point particle", "velocity", and "classical trajectory", were unsuccessful [5,6]. This list should also include the so-called Bohmian mechanics [7-9], which cannot be considered a completely classical formulation of quantum mechanics because the velocity field, in which a motion of individual particles is calculated, is the solution of the Schrödinger equation. For this reason, Bohmian mechanics should be called Bohmian kinematics, while the corresponding dynamics is still described by the Schrödinger equation.

Quantum mechanics is associated with classical mechanics through the Hamilton-Jacobi theory [4]. If we represent the wave function in the form of $\psi = \sqrt{\rho}\exp(iS/\hbar)$ and separate the Schrödinger equation into real and imaginary parts, one arrives at both the continuity equation (3) and the following equation:

$$\frac{\partial S}{\partial t} + \frac{1}{2m}|\nabla S|^2 + U - \frac{\hbar^2}{2m}\frac{\nabla^2\sqrt{\rho}}{\sqrt{\rho}} = 0 \qquad (7)$$

which, formally, is the Hamilton-Jacobi equation for a classical particle moving in a potential field

$$U_{ef}(\mathbf{r},t) = U - \frac{\hbar^2}{2m}\frac{\nabla^2\sqrt{\rho}}{\sqrt{\rho}} \qquad (8)$$

where

$$U_q(\mathbf{r},t) = -\frac{\hbar^2}{2m}\frac{\nabla^2\sqrt{\rho}}{\sqrt{\rho}} \qquad (9)$$

is so-called quantum potential associated with the wave properties of the quantum particle.

The motion of a quantum particle in an external potential field $U(\mathbf{r})$, at least formally, is equivalent to the motion of a classical particle in a potential field (8). For this reason, one can say that the Schrödinger ensemble is a quantum Hamilton-Jacobi ensemble, which we call the Hamilton-Jacobi-Schrödinger ensemble.

The Schrödinger equation in the form of Eqs. (3) and (7) is used to justify the transition from quantum to classical mechanics in the limit $\hbar \to 0$: in this limit, the quantum mechanics becomes the Hamilton-Jacobi theory for a classical particle.

One of interpretations of quantum mechanics, Bohmian mechanics, is based on this analogy.

Using (7) and (8), one can formally write Newton's second law (1) for a quantum particle as

$$m\ddot{\mathbf{r}} = -\nabla U_{ef} \qquad (10)$$

Thus, the motion of quantum particle, at least formally, can be calculated within the limits of classical mechanics if the effective potential field (8) is known.

However, the difficulty of this approach is that the quantum potential (9) is not a predetermined function of the coordinates, as in classical mechanics. Instead, it depends on the probability density (the density of the ensemble) $\rho(\mathbf{r},t)$, which can be found using the solution of the Schrödinger equation.

If one considers equations (3), (4), and (7) as a hydrodynamic description of the Hamilton-Jacobi-Schrodinger ensemble (Madelung fluid), in this case, the quantum potential (9) plays the role of "pressure". A Madelung fluid is compressible one, but the "pressure" does not depend on the density, as it does for a classical compressible fluid; it depends on the second derivatives of the density with respect to the coordinates. Thus, even such a semi-classical hydrodynamic model has fundamental difficulties, both from the perspective of the classical interpretation of quantum mechanics and from the numerical (hydrodynamic) modeling of quantum particle motion.

In this paper, we show that there is another hydrodynamic formulation of quantum mechanics in which there are no such problems and that is close to the hydrodynamic description of flows of classical inviscid ideal gases.

## II. VARIATIONAL PRINCIPLE IN CLASSICAL AND QUANTUM MECHANICS

The variational principles in physics play an important role. On the one hand, they are a formal way to derive the fundamental laws of nature [1,10,11], and, on the other hand, they are part of a philosophical principle that shows that Nature is arranged rationally and "spends a minimal effort" in its development.

The equations of motion of a classical system of point particles are derived from the least action principle [1]:

$$\delta S = 0 \tag{11}$$

under the condition

$$\delta q(t) = \delta q(t_0) \tag{12}$$

where

$$S = S_0 + \int_{t_0}^{t} L(q,\dot{q},t)dt \tag{13}$$

$S_0 = S(t_0)$, $t_0$ are the constants, $L(q,\dot{q},t)$ is the Lagrange function, and $q(t)$ are the generalized coordinates of the system.

For one point particle, moving in a potential field $U$,

$$L = \frac{m\mathbf{v}^2}{2} - U(\mathbf{r}) \tag{14}$$

$$S = S_0 + \int_{t_0}^{t} \left( \frac{m\mathbf{v}^2}{2} - U(\mathbf{r}) \right) dt \tag{15}$$

while, for the true trajectories of the particle,

$$S \to \min \tag{16}$$

(or, more precisely, it tends to a steady value) for any instant $t$, under conditions (12) and constant $S_0, t_0$.

The expression (15) can be rewritten in the form

$$\int_{t_0}^{t}\left(\frac{dS}{dt}-\frac{m\mathbf{v}^2}{2}+U(\mathbf{r})\right)dt=0 \tag{17}$$

or, using the $\delta$-function, as

$$\int_{t_0}^{t}\int\delta(\mathbf{r}-\mathbf{r}_s(t))\left(\frac{dS}{dt}-\frac{m\mathbf{v}^2}{2}+U(\mathbf{r})\right)d\mathbf{r}dt=0 \tag{18}$$

where $\mathbf{r}_s(t)$ is a motion law of the particle and the integral with respect to $\mathbf{r}$ is taken over the entire space (the entire configuration space for the system of particles).

Let us consider the Hamilton-Jacobi ensemble consisting of a set of identical non-interacting particles with different initial conditions. The Hamilton-Jacobi ensemble corresponding to a single particle can be represented as a compressible fluid (gas), where the flow is described by the velocity field (4).

In this case, we should transfer from an individual (Lagrangian) description, which is used for a single particle, to a continual (Euler) description [12]. As a result, one obtains

$$\frac{dS}{dt}=\frac{\partial S}{\partial t}+(\mathbf{v}\nabla)S \tag{19}$$

where, in accordance with (4),

$$\mathbf{v}=\frac{1}{m}\nabla S \tag{20}$$

Summing (18) over all of the particles in the Hamilton-Jacobi ensemble, one obtains

$$\int_{t_0}^{t}\int\sum_{s}\delta(\mathbf{r}-\mathbf{r}_s(t))\left(\frac{\partial S}{\partial t}+\frac{1}{2m}|\nabla S|^2+U(\mathbf{r})\right)d\mathbf{r}dt=0 \tag{21}$$

Let us turn to a continuous distribution of particles in the Hamilton-Jacobi ensemble over space. For this purpose, one introduces the density of particles over space (density of ensemble):

$$\rho(\mathbf{r},t)=\sum_{s}\delta(\mathbf{r}-\mathbf{r}_s(t)) \tag{22}$$

Evidently,

$$\int_{\Omega}\rho(\mathbf{r},t)d\mathbf{r}=N$$

is the number of particles in a volume $\Omega$.

It is convenient to redetermine the density (22) using

$$\rho(\mathbf{r},t)=\frac{1}{N}\sum_{s}\delta(\mathbf{r}-\mathbf{r}_s(t))$$

In this case, $\rho(\mathbf{r},t)$ has a normalization

$$\int_\Omega \rho(\mathbf{r},t)d\mathbf{r} = 1$$

and can be considered a probability density to find the particle of the Hamilton-Jacobi ensemble at a given point.

Eq. (21) then takes the following form:

$$\int_{t_0}^{t}\int \rho(\mathbf{r},t)\left(\frac{\partial S}{\partial t} + \frac{1}{2m}|\nabla S|^2 + U(\mathbf{r})\right)d\mathbf{r}dt = 0 \quad (23)$$

Thus, the true motion of the Hamilton-Jacobi ensemble (and thus the true motion of individual particles in this ensemble) corresponds to the functions $\rho(\mathbf{r},t)$ and $S(\mathbf{r},t)$ that satisfy the condition (23) and the conditions (11) and (12) simultaneously.

We assume the functions $\rho(\mathbf{r},t)$ and $S(\mathbf{r},t)$ are independent. By varying the expression in Eq. (23) with respect to $\rho(\mathbf{r},t)$ and equating the variation to zero, one obtains the Hamilton-Jacobi equation (5); by varying the expression in Eq. (23) with respect to $S(\mathbf{r},t)$ and equating the variation to zero, one obtains the continuity equation for the Hamilton-Jacobi ensemble (3), (4). Thus, we arrive at the variational principle of Hamilton-Jacobi theory in classical mechanics.

Let us now consider quantum mechanics. The Schrödinger equation can be written in the form of (3), (4), and (7), which is formally identical to the classical Hamilton-Jacobi theory for a particle in the effective potential field (8), which depends on the density of the ensemble $\rho$. Generalizing the expression (23) to quantum particles, one can write

$$\int_{t_0}^{t}\int \rho(\mathbf{r},t)\left(\frac{\partial S}{\partial t} + \frac{1}{2m}|\nabla S|^2 + U_{ef}(\mathbf{r})\right)d\mathbf{r}dt = 0 \quad (24)$$

This generalization is based on the formal similarity of equations (5) and (7). The quantum potential (9) can be written as

$$U_q = \frac{\hbar^2}{2m}\left(\frac{1}{4\rho^2}|\nabla\rho|^2 - \frac{1}{2\rho}\nabla^2\rho\right) \quad (25)$$

By substituting Eq. (8) in Eq. (24) and considering Eq. (25), after simple transformations, one obtains

$$\int_{t_0}^{t}\int \rho(\mathbf{r},t)\left(\frac{\partial S}{\partial t} + \frac{1}{2m}|\nabla S|^2 + U(\mathbf{r}) + \frac{\hbar^2}{8m}|\nabla \ln \rho|^2\right)d\mathbf{r}dt - \frac{\hbar^2}{4m}\int_{t_0}^{t}\oint(d\boldsymbol{\sigma}\nabla)\rho\, dt = 0 \quad (26)$$

where the integral $\oint(d\boldsymbol{\sigma}\nabla)\rho$ is taken over the surface bounding the space region in which the Hamilton-Jacobi-Schrodinger ensemble moves and $d\boldsymbol{\sigma}$ is the vector element of area of this surface.

We assume that, at the boundary of the region occupied by Hamilton-Jacobi-Schrödinger ensemble,

$$(d\boldsymbol{\sigma}\nabla)\rho = 0$$

This condition can always be satisfied because one can always consider a region of space to be much larger than the size of the Hamilton-Jacobi-Schrödinger ensemble, such that the density of the ensemble at the region boundary is a constant or even equal to zero.

As a consequence, one obtains

$$\int_{t_0}^{t}\int \rho(\mathbf{r},t)\left(\frac{\partial S}{\partial t} + \frac{1}{2m}|\nabla S|^2 + U(\mathbf{r}) + \frac{\hbar^2}{8m}|\nabla \ln \rho|^2\right)d\mathbf{r}dt = 0 \tag{27}$$

By direct calculation, it is easy to check that the independent variation of Eq. (27) with respect to $S(\mathbf{r},t)$ leads to the continuity equations (3) and (4), while the variation with respect to $\rho(\mathbf{r},t)$ leads to the Hamilton-Jacobi equation (7) with the effective potential energy (8). Together, these equations form the Schrödinger equation.

Thus, the transition from the classical variational principle for Eq. (23) to the quantum-mechanical variational principle for Eq. (24) is justified based on the formal similarity of equations (5) and (7), although, in quantum mechanics, the effective potential energy (8) is varied by varying the density $\rho(\mathbf{r},t)$.

The expression (25) shows that the quantum potential $U_q$ can be written as

$$U_q = U_q' + U_q'' \tag{28}$$

where

$$U_q' = \frac{\hbar^2}{8m}|\nabla \ln \rho|^2 \tag{29}$$

$$U_q'' = -\frac{\hbar^2}{4m}\frac{1}{\rho}\nabla^2\rho \tag{30}$$

It is interesting to note that both components (29) and (30) of the quantum potential enter into the Hamilton-Jacobi equation (7) for a quantum particle, while only the component $U_q'$ of the quantum potential appears in the integral variational principle (27). Based on this analysis, we conclude that precisely $U_q'$ (29) may be called the quantum potential, while the component $U_q''$ appears in the equation (7) due to varying the potential (29) with respect to density. Thus, from the point of view of the variational principle (27), the motion of the quantum particle is equivalent to the motion of a classical particle in a potential field of

$$U_{ef} = U + \frac{\hbar^2}{8m}|\nabla \ln \rho|^2 \tag{31}$$

This effective potential differs from the potential (8) based on the formal similarity of the Schrödinger equation in the form of (7) and the Hamilton-Jacobi equation (5). As we show below, precisely the potential (29) is responsible for the unusual (non-classical) behavior of quantum particles.

### III. HYDRODYNAMICS OF THE HAMILTON-JACOBI-SCHRÖDINGER ENSEMBLE

#### A. Particle motion in a quickly oscillating field

Let us consider the motion of a classical particle in an external (slowly changing) potential field $U(\mathbf{r},t)$, on which the following quickly oscillating force simultaneously acts:

$$\mathbf{f}(\mathbf{r},t) = \mathbf{f}_c(\mathbf{r},t)\cos\omega t + \mathbf{f}_s(\mathbf{r},t)\sin\omega t \qquad (32)$$

where $\omega$ is the frequency and $\mathbf{f}_c(\mathbf{r},t)$, $\mathbf{f}_s(\mathbf{r},t)$ are the vectors, which depend on the coordinates and are weakly dependent on time. The frequency $\omega$ satisfies the condition $\omega \gg 1/T$, where $T$ is the characteristic time of the particle motion in an external field $U(\mathbf{r},t)$ at $\mathbf{f}_c(\mathbf{r},t) = \mathbf{f}_s(\mathbf{r},t) = 0$. The weak dependence of the vectors $\mathbf{f}_c(\mathbf{r},t)$, $\mathbf{f}_s(\mathbf{r},t)$ on the time means that the characteristic time of their change is much bigger than $1/\omega$.

Under the action of the external force (32), a particle performs a complex motion that consists of the average motion along a smooth trajectory $\mathbf{R}(t) = \langle \mathbf{r}(t) \rangle$ and the fast oscillations with a frequency $\omega$ around it.

It is well known [1] that, averaged over the oscillation, the motion of the particle is described by the equation

$$m\ddot{\mathbf{R}} = -\nabla U_{ef} \qquad (33)$$

where

$$U_{ef}(\mathbf{r},t) = U(\mathbf{r},t) + \frac{1}{4m\omega^2}\left(|\mathbf{f}_c(\mathbf{r},t)|^2 + |\mathbf{f}_s(\mathbf{r},t)|^2\right) \qquad (34)$$

Thus, the action of the quickly oscillating force (32) results in the creation of an additional potential energy of

$$U_p = \frac{1}{4m\omega^2}\left(|\mathbf{f}_c(\mathbf{r},t)|^2 + |\mathbf{f}_s(\mathbf{r},t)|^2\right) \qquad (35)$$

which is simply the kinetic energy of the oscillatory motion.

For example, when a charged particle moves in the field of an electromagnetic wave, a quickly oscillating force (32) is caused by an electric field: $\mathbf{f}(\mathbf{r},t) = q\mathbf{E}_0(\mathbf{r},t)\exp(i\omega t)$, where $q$ is the electric charge of the particle and $\mathbf{E}_0(\mathbf{r},t)$ is the slowly varying amplitude of the electric field. In this case, the additional potential energy (35) takes the form

$$U_p = \frac{q^2}{4m\omega^2}|\mathbf{E}_0(\mathbf{r},t)|^2 \qquad (36)$$

and is called the ponderomotive potential [13].

## B. Euler equation

By comparing Eqs. (34) and (31), one can conclude that the quantum potential (29) can be represented, at least formally, as an additional potential energy (35) that arises as a result of the action of the quickly oscillating force:

$$\mathbf{f}(\mathbf{r},t) = \frac{1}{\sqrt{2}}\hbar\omega\nabla\ln\rho\cos\omega t \qquad (37)$$

Thus, the component of the quantum potential (29) can be explained within the limits of "classical" mechanics if one assumes that a quickly oscillating force (37) other than an external potential $U(\mathbf{r},t)$ acts on a point particle. By definition, this force has a very high frequency $\omega$ and, correspondingly, a large amplitude $\sim\omega$.

Of course, this force is not a true classical one because it depends on the density of the Hamilton-Jacobi-Schrödinger ensemble. The interpretation of this force is considered below. The analysis above shows that the quantum particle can be considered a classical particle moving in an external field $U(\mathbf{r},t)$, on which an additional quickly oscillating force (37) acts. Turning to the Hamilton-Jacobi-Schrodinger ensemble, it is easy to see that the quickly oscillating force per unit volume of the Hamilton-Jacobi-Schrödinger ensemble is

$$\rho\mathbf{f}(\mathbf{r},t) = \frac{1}{\sqrt{2}}\hbar\omega\cos\omega t\,\nabla\rho \qquad (38)$$

In the continual description, one can write the equation of motion for the Hamilton-Jacobi-Schrödinger ensemble as

$$m\rho_r\left(\frac{\partial\mathbf{v}_r}{\partial t} + (\mathbf{v}_r\nabla)\mathbf{v}_r\right) = -\rho_r\nabla U + \frac{1}{\sqrt{2}}\hbar\omega\cos\omega t\nabla\rho_r \qquad (39)$$

which is the hydrodynamic Euler equation.

This equation should be solved together with the continuity equation:

$$\frac{\partial\rho_r}{\partial t} + \operatorname{div}(\rho_r\mathbf{v}_r) = 0 \qquad (40)$$

Here and below, the subscript "$r$" refers to the true (microscopic) parameters of the Hamilton-Jacobi-Schrödinger ensemble, including the quick oscillations, while the parameters without the subscript refer to the average motion of the ensemble.

The last term on the right-hand side of Eq. (39) can be written as ($-\nabla p$), where

$$p = -\frac{1}{\sqrt{2}} \hbar \omega \rho_r \cos \omega t \qquad (41)$$

can be interpreted as the "pressure" in the Hamilton-Jacobi-Schrödinger ensemble, which is a quickly oscillating function of time. The quantum Hamilton-Jacobi-Schrödinger ensemble is an inviscid gas with internal pressure (41). However, this pressure is not the conventional pressure that occurs in a classical gas because it is sign-alternating and therefore cannot be explained by the classical kinetic model [14].

Formally, the relation (41) can be considered the equation of state of a classical ideal gas $p = kT\rho$. In this case, the "temperature" of the Hamilton-Jacobi-Schrödinger gas is determined by the expression

$$T = -\frac{1}{\sqrt{2}} (\hbar \omega / k) \cos \omega t \qquad (42)$$

which is also sign-alternating; for this reason, it cannot be considered as an "average kinetic energy" of the random motion of particles.

The velocity $\mathbf{v}_r$ that enters into the Euler equation (39) and the continuity equation (40) is the instantaneous velocity of the particles in the ensemble, while the velocity defined by Eq. (4) is an average over the fast oscillations of the particles' velocities.

At constant $\omega$, the Euler equation (39) has a solution in the form of the potential flow of the Hamilton-Jacobi-Schrödinger ensemble:

$$\mathbf{v}_r = \frac{1}{m} \nabla S_r \qquad (43)$$

where the function $S_r(\mathbf{r},t)$ is different than the action defined by the Schrödinger equation (7) and satisfies the Hamilton-Jacobi equation

$$\frac{\partial S_r}{\partial t} + \frac{1}{2m} |\nabla S_r|^2 + U - \frac{1}{\sqrt{2}} \hbar \omega \cos \omega t \ln \rho_r = 0 \qquad (44)$$

The density $\rho_r$ satisfies the continuity equation (40) with the velocity (43).

Equation (44) differs from the Hamilton-Jacobi equation (5) for a classical particle in that it contains a quickly oscillating potential that depends on the density of the Hamilton-Jacobi-Schrödinger ensemble.

Formally, equation (44) can be considered an ordinary Hamilton-Jacobi equation for a classical particle moving in a potential field

$$U_{ref} = U - \frac{1}{\sqrt{2}} \hbar \omega \cos \omega t \ln \rho_r \tag{45}$$

The component

$$U_{rq} = -\frac{1}{\sqrt{2}} \hbar \omega \cos \omega t \ln \rho_r \tag{46}$$

can be called a true quantum potential in contrast to $U'_q$ in Eq. (29), which is the result of averaging the particle motion over the fast oscillations.

Equations (39) and (40) are formally the hydrodynamic equations of an inviscid ideal gas with a quickly oscillating sign-alternating temperature (pressure) and can easily be solved numerically using the conventional methods of classical computational fluid dynamics.

Passing from the Hamilton-Jacobi theory to Newton's description of a quantum particle, one can formally write the Newton's law as

$$m \frac{d\mathbf{v}_r}{dt} = -\nabla U_{ref} \tag{47}$$

## IV. DIRECT DERIVATION OF THE SCHRÖDINGER EQUATION

### A. Particle in a potential field

In the previous section, based on rather general but non-rigorous reasoning, we came to equation (39), which, together with the continuity equation (40), should be equivalent to the Schrödinger equation because it gives a detailed description of the motion of a quantum particle (more precisely, the Hamilton-Jacobi-Schrödinger ensemble); however, averaging these equations over the fast oscillations should lead to the Schrödinger equation. Let us show that the motion of the Hamilton-Jacobi-Schrödinger ensemble, as described by equations (39) and (40), is actually described by the Schrödinger equation when averaged over the fast oscillations.

This analysis is more simply performed using the Hamilton-Jacobi equation (44) and the continuity equation (40) with the velocity in Eq. (43).

Let us seek the solution of Eqs. (40), (43), (44) in the form

$$\rho_r = \rho + \zeta, \; S_r = S + \sigma \tag{48}$$

where $\rho$ and $S$ are slowly varying functions with a characteristic time scale $T \gg 1/\omega$ and $\zeta$ and $\sigma$ are the quickly oscillating functions with frequency $\omega$ that satisfy the conditions

$$\langle \zeta \rangle = 0; \; \langle \sigma \rangle = 0 \tag{49}$$

Here, $\langle...\rangle$ denotes averaging over the fast oscillations.

By substituting Eq. (48) into Eqs. (40), (43) and (44), one obtains

$$\frac{\partial S}{\partial t}+\frac{\partial \sigma}{\partial t}+\frac{1}{2m}|\nabla S|^2+\frac{1}{m}(\nabla S \nabla \sigma)+\frac{1}{2m}|\nabla \sigma|^2+U- \\ -\frac{1}{\sqrt{2}}\hbar\omega\cos\omega t \ln\rho-\frac{1}{\sqrt{2}}\hbar\omega(\zeta/\rho)\cos\omega t = 0 \tag{50}$$

$$\frac{\partial \rho}{\partial t}+\frac{\partial \zeta}{\partial t}+\mathrm{div}\left(\frac{1}{m}\rho\nabla S+\frac{1}{m}\zeta\nabla S+\frac{1}{m}\rho\nabla\sigma+\frac{1}{m}\zeta\nabla\sigma\right)=0 \tag{51}$$

When writing equation (50), one assumes that $|\zeta|/\rho \ll 1$, so $\ln(1+\zeta/\rho)\approx \zeta/\rho$. This condition is proven below.

Averaging equations (50) and (51) over the fast oscillations while considering Eq. (49) allows the separation of the fast and slow components. As a result, one obtains

$$\frac{\partial S}{\partial t}+\frac{1}{2m}|\nabla S|^2+\frac{1}{2m}\langle|\nabla\sigma|^2\rangle+U-\frac{1}{\sqrt{2}\rho}\hbar\omega\langle\zeta\cos\omega t\rangle=0 \tag{52}$$

$$\frac{\partial \rho}{\partial t}+\mathrm{div}\left(\frac{1}{m}\rho\nabla S+\frac{1}{m}\langle\zeta\nabla\sigma\rangle\right)=0 \tag{53}$$

$$\frac{\partial \sigma}{\partial t}+\frac{1}{m}(\nabla S \nabla \sigma)+\frac{1}{2m}\left(|\nabla\sigma|^2-\langle|\nabla\sigma|^2\rangle\right)-\frac{1}{\sqrt{2}}\hbar\omega\cos\omega t \ln\rho+ \\ +\frac{1}{\sqrt{2}\rho}\hbar\omega(\langle\zeta\cos\omega t\rangle-\zeta\cos\omega t)=0 \tag{54}$$

$$\frac{\partial \zeta}{\partial t}+\mathrm{div}\left(\frac{1}{m}\zeta\nabla S+\frac{1}{m}\rho\nabla\sigma+\frac{1}{m}(\zeta\nabla\sigma-\langle\zeta\nabla\sigma\rangle)\right)=0 \tag{55}$$

By considering oscillations $\zeta$ and $\sigma$ small, we can restrict our consideration to the linear approximation, in which equations (54) and (55) can be written as

$$\frac{\partial \sigma}{\partial t}+\frac{1}{m}(\nabla S \nabla \sigma)-\frac{1}{\sqrt{2}}\hbar\omega\cos\omega t \ln\rho=0 \tag{56}$$

$$\frac{\partial \zeta}{\partial t}+\mathrm{div}\left(\frac{1}{m}\zeta\nabla S+\frac{1}{m}\rho\nabla\sigma\right)=0 \tag{57}$$

We show below that the convective terms $\frac{1}{m}(\nabla S \nabla \sigma)$ and $\frac{1}{m}\zeta\nabla S$, which are related to the average flow of the Hamilton-Jacobi-Schrödinger ensemble, are much less than the other terms in equations (56) and (57).

Thus, the terms in Eqs. (56) and (57) describing the convective transport, which is associated with the average flow of Hamilton-Jacobi-Schrödinger ensemble, can be neglected. As a result, equations (56) and (57) take the form

$$\frac{\partial \sigma}{\partial t} - \frac{1}{\sqrt{2}} \hbar \omega \cos \omega t \ln \rho = 0 \tag{58}$$

$$\frac{\partial \zeta}{\partial t} + \mathrm{div}\left(\frac{1}{m} \rho \nabla \sigma\right) = 0 \tag{59}$$

Equations (58) and (59) have the solutions

$$\sigma = \frac{1}{\sqrt{2}} \hbar \sin \omega t \ln \rho \tag{60}$$

$$\zeta = \frac{1}{\sqrt{2}} \frac{\hbar}{m\omega} \cos \omega t \, \nabla^2 \rho \tag{61}$$

Let us compare the convective term $\frac{1}{m}(\nabla S \nabla \sigma)$ with a quickly oscillating component $\frac{1}{\sqrt{2}} \hbar \omega \cos \omega t \ln \rho$ in equation (56) using the solution (60). Because the oscillations of these terms are sign-alternating with zero mean values, it makes sense to compare their root-mean square values.

We have the estimations

$$\frac{1}{m} \sqrt{\langle |(\nabla S \nabla \sigma)|^2 \rangle} = \frac{\hbar}{2m} |(\nabla S \nabla \ln \rho)| \sim \frac{\hbar V}{2L} |\ln \rho| \sim \frac{\hbar}{2T} |\ln \rho|$$

$$\langle \left(\frac{1}{\sqrt{2}} \hbar \omega \cos \omega t \ln \rho\right)^2 \rangle = \frac{1}{2} \hbar \omega |\ln \rho|$$

where $V \sim \frac{1}{m}|\nabla S|$ is the characteristic velocity of the mean flow of the Hamilton-Jacobi-Schrödinger ensemble, $L$ is the characteristic spatial scale of the mean flow of the Hamilton-Jacobi-Schrödinger ensemble, and $T = L/V$ is the characteristic time of the change of the parameters of the mean flow of Hamilton-Jacobi-Schrödinger ensemble.

Because we consider very fast oscillations satisfying the condition $1/T \ll \omega$, the estimation above shows that the term $\frac{1}{m}(\nabla S \nabla \sigma)$, in mean, is much less than the quickly oscillating term $\frac{1}{\sqrt{2}} \hbar \omega \cos \omega t \ln \rho$ and can be discarded, which was done in equation (58).

Similarly, one can compare the components $\zeta \nabla S$ and $\rho \nabla \sigma$ in equation (57) using the solutions of Eqs (60) and (61).

One obtains

$$\sqrt{\langle \zeta^2 \rangle} |\nabla S| = \frac{\hbar}{2m\omega} |\nabla^2 \rho| |\nabla S| \sim \frac{\hbar V}{2\omega L^2} \rho = \frac{1}{\omega T} \frac{\hbar}{2L} \rho$$

$$\rho\sqrt{\langle|\nabla\sigma|^2\rangle} = \frac{\hbar}{2}|\nabla\rho| \sim \frac{\hbar}{2L}\rho$$

Given that $\frac{1}{\omega T} \ll 1$ in the case considered, one concludes that the term $\zeta\nabla S$, in mean, is much less than the term $\rho\nabla\sigma$ and can be neglected, which was done in equation (59). Let us estimate the ratio $|\zeta|/\rho$:

$$\sqrt{\langle\zeta^2\rangle}/\rho = \frac{1}{2}\frac{\hbar}{m\omega\rho}|\nabla^2\rho| \sim \frac{1}{2}\frac{\hbar}{m\omega L^2} = \frac{1}{2}\frac{\hbar}{mVL\omega T} \sim \frac{1}{2}\frac{\hbar}{S_m}\frac{1}{\omega T}$$

where $S_m = mVL$ is the scale of action for the mean flow of the Hamilton-Jacobi-Schrödinger ensemble. For quantum particles $S_m \geq \hbar$, and thus one concludes that the condition $|\zeta|/\rho \ll 1$ is satisfied in mean and that the higher the frequency $\omega$ in comparison with $1/T$ the more precisely it is satisfied.

Using the solutions (60) and (61), it is easy to obtain

$$\langle\zeta\nabla\sigma\rangle = 0 \tag{62}$$

$$\frac{1}{2m}\langle|\nabla\sigma|^2\rangle = \frac{\hbar^2}{8m}\frac{|\nabla\rho|^2}{\rho^2} \tag{63}$$

$$\frac{1}{\sqrt{2}\rho}\hbar\omega\langle\zeta\cos\omega t\rangle = \frac{\hbar^2}{4m}\frac{\nabla^2\rho}{\rho} \tag{64}$$

Substituting Eqs. (62) through (64) into equations (52) and (53) for the mean motion of the Hamilton-Jacobi-Schrödinger ensemble gives

$$\frac{\partial S}{\partial t} + \frac{1}{2m}|\nabla S|^2 + U + \frac{\hbar^2}{8m}\frac{|\nabla\rho|^2}{\rho^2} - \frac{\hbar^2}{4m}\frac{\nabla^2\rho}{\rho} = 0 \tag{65}$$

$$\frac{\partial\rho}{\partial t} + \text{div}\left(\frac{1}{m}\rho\nabla S\right) = 0 \tag{66}$$

This set of equations coincides with equations (3), (4) and (7) of the Hamilton-Jacobi theory, with the quantum potential (25) and, consequently, with the Schrödinger equation for the wave function $\psi = \sqrt{\rho}\exp(iS/\hbar)$.

We have proven that equations (39) and (40) are equivalent to the Schrödinger equation by reasoning from the Hamilton-Jacobi equation (44). The same result can be obtained by considering the Euler equation (39) directly and writing the velocity of the Hamilton-Jacobi ensemble in the form $\mathbf{v}_r = \mathbf{v} + \mathbf{w}$, where $\mathbf{w}$ is the quickly oscillating component of the velocity satisfying the condition $\langle\mathbf{w}\rangle = 0$ and $\mathbf{v}$ is the mean velocity of flow of the Hamilton-Jacobi-Schrödinger ensemble.

## B. Charged particle in an electromagnetic field

We considered the particle in an external potential field and demonstrated that the motion of the Hamilton-Jacobi-Schrödinger ensemble, being described by Eqs. (39) and (40), after averaging over the fast oscillations, is described by the Schrödinger equation. Now, we will show that this statement is also true for a non-relativistic spinless particle moving in an external electromagnetic field.

Let the Lorentz force $\mathbf{F} = q\left(\mathbf{E} + \frac{1}{c}[\mathbf{vH}]\right)$ act on a charged particle, where $q$ is the particle charge, $\mathbf{E} = -\nabla\varphi - \frac{1}{c}\frac{\partial \mathbf{A}}{\partial t}$, $\mathbf{H} = \text{rot}\mathbf{A}$ are the external electric and magnetic fields, and $\varphi, \mathbf{A}$ are the scalar and vector potentials of the electromagnetic field.

The natural generalization of Eq. (39) is as follows:

$$m\rho_r\left(\frac{\partial \mathbf{v}_r}{\partial t} + (\mathbf{v}_r\nabla)\mathbf{v}_r\right) = \rho_r\, q\left(\mathbf{E} + \frac{1}{c}[\mathbf{v}_r\mathbf{H}]\right) + \frac{1}{\sqrt{2}}\hbar\omega\cos\omega t\,\nabla\rho_r \qquad (67)$$

Let us assume that the vector and scalar potentials of the external electromagnetic field, $\mathbf{A}$ and $\varphi$, are slowly varying functions in that the characteristic time of their changing is $T \ll 1/\omega$. Substituting the electric and magnetic fields, expressed in terms of the scalar and vector potentials of the electromagnetic field, into Eq. (67) gives

$$\frac{\partial}{\partial t}\left(m\mathbf{v}_r + \frac{q}{c}\mathbf{A}\right) + m(\mathbf{v}_r\nabla)\mathbf{v}_r - \frac{q}{c}[\mathbf{v}_r\text{rot}\mathbf{A}] = -q\nabla\varphi + \frac{1}{\sqrt{2}}\hbar\omega\cos\omega t\,\nabla\ln\rho_r \qquad (68)$$

Given that

$$(\mathbf{v}_r\nabla)\mathbf{v}_r = \frac{1}{2}\nabla\mathbf{v}_r^2 - [\mathbf{v}_r\,\text{rot}\,\mathbf{v}_r]$$

we rewrite Eq. (68) in the form

$$\frac{\partial}{\partial t}\left(m\mathbf{v}_r + \frac{q}{c}\mathbf{A}\right) + \nabla\left(\frac{m\mathbf{v}_r^2}{2} + q\varphi\right) = [\mathbf{v}_r\,\text{rot}\left(m\mathbf{v}_r + \frac{q}{c}\mathbf{A}\right)] + \frac{1}{\sqrt{2}}\hbar\omega\cos\omega t\,\nabla\ln\rho_r \qquad (69)$$

Equation (69) has a solution in the form of

$$m\mathbf{v}_r + \frac{q}{c}\mathbf{A} = \nabla S_r \qquad (70)$$

where the function $S_r$ satisfies the equation

$$\frac{\partial S_r}{\partial t} + \frac{m}{2}\mathbf{v}_r^2 + q\varphi - \frac{1}{\sqrt{2}}\hbar\omega\cos\omega t\,\ln\rho_r = 0 \qquad (71)$$

which, by considering Eq. (70), is the Hamilton-Jacobi equation for a non-relativistic particle in an external electromagnetic field with an additional oscillating potential (41):

$$\frac{\partial S_r}{\partial t} + \frac{1}{2m}\left(\nabla S_r - \frac{q}{c}\mathbf{A}\right)^2 + q\varphi - \frac{1}{\sqrt{2}}\hbar\omega\cos\omega t \ln\rho_r = 0 \qquad (72)$$

Correspondingly, the continuity equation (40) takes the form

$$\frac{\partial \rho_r}{\partial t} + \text{div}\left[\rho_r \frac{1}{m}\left(\nabla S_r - \frac{q}{c}\mathbf{A}\right)\right] = 0 \qquad (73)$$

As in the previous section, we seek the solution of equations (72) and (73) in the form of (48) and (49).

Substituting Eq. (48) into Eqs. (72) and (73) gives

$$\frac{\partial S}{\partial t} + \frac{\partial \sigma}{\partial t} + \frac{1}{2m}\left(\nabla S - \frac{q}{c}\mathbf{A}\right)^2 + \frac{1}{m}\left(\nabla S - \frac{q}{c}\mathbf{A}\right)\cdot\nabla\sigma + \frac{1}{2m}|\nabla\sigma|^2 + U - \\ -\frac{1}{\sqrt{2}}\hbar\omega\cos\omega t \ln\rho - \frac{1}{\sqrt{2}}\hbar\omega(\zeta/\rho)\cos\omega t = 0 \qquad (74)$$

$$\frac{\partial\rho}{\partial t} + \frac{\partial\zeta}{\partial t} + \text{div}\left(\frac{1}{m}\rho\left(\nabla S - \frac{q}{c}\mathbf{A}\right) + \frac{1}{m}\zeta\left(\nabla S - \frac{q}{c}\mathbf{A}\right) + \frac{1}{m}\rho\nabla\sigma + \frac{1}{m}\zeta\nabla\sigma\right) = 0 \qquad (75)$$

Here, as above, it is assumed that, on average, $|\zeta|/\rho \ll 1$, so $\ln(1+\zeta/\rho) \approx \zeta/\rho$.

Averaging equations (74) and (75) over the fast oscillations, while considering Eq. (49) allows the separation of the fast and the slow components. As a result, one obtains

$$\frac{\partial S}{\partial t} + \frac{1}{2m}\left(\nabla S - \frac{q}{c}\mathbf{A}\right)^2 + \frac{1}{2m}\langle|\nabla\sigma|^2\rangle + U - \frac{1}{\sqrt{2}\rho}\hbar\omega\langle\zeta\cos\omega t\rangle = 0 \qquad (76)$$

$$\frac{\partial\rho}{\partial t} + \text{div}\left(\frac{1}{m}\rho\left(\nabla S - \frac{q}{c}\mathbf{A}\right) + \frac{1}{m}\langle\zeta\nabla\sigma\rangle\right) = 0 \qquad (77)$$

$$\frac{\partial\sigma}{\partial t} + \frac{1}{m}\left(\nabla S - \frac{q}{c}\mathbf{A}\right)\cdot\nabla\sigma + \frac{1}{2m}\left(|\nabla\sigma|^2 - \langle|\nabla\sigma|^2\rangle\right) - \frac{1}{\sqrt{2}}\hbar\omega\cos\omega t \ln\rho + \\ + \frac{1}{\sqrt{2}\rho}\hbar\omega(\langle\zeta\cos\omega t\rangle - \zeta\cos\omega t) = 0 \qquad (78)$$

$$\frac{\partial\zeta}{\partial t} + \text{div}\left(\frac{1}{m}\zeta\left(\nabla S - \frac{q}{c}\mathbf{A}\right) + \frac{1}{m}\rho\nabla\sigma + \frac{1}{m}(\zeta\nabla\sigma - \langle\zeta\nabla\sigma\rangle)\right) = 0 \qquad (79)$$

Assuming that the oscillations $\zeta$ and $\sigma$ are small, equations (78) and (79) can be written in linear approximation in the following form:

$$\frac{\partial\sigma}{\partial t} + \frac{1}{m}\left(\nabla S - \frac{q}{c}\mathbf{A}\right)\cdot\nabla\sigma - \frac{1}{\sqrt{2}}\hbar\omega\cos\omega t \ln\rho = 0 \qquad (80)$$

$$\frac{\partial \zeta}{\partial t} + \mathrm{div}\left(\frac{1}{m}\zeta\left(\nabla S - \frac{q}{c}\mathbf{A}\right) + \frac{1}{m}\rho\nabla\sigma\right) = 0 \tag{81}$$

By analogy with the previous section, it can be shown that the terms in equations (80) and (81) describing the convective transport with the mean flow of the Hamilton-Jacobi-Schrödinger ensemble are, on average, significantly less than the other terms in these equations and can be neglected. As a result, Eqs. (80) and (81) take the form

$$\frac{\partial \sigma}{\partial t} - \frac{1}{\sqrt{2}}\hbar\omega\cos\omega t \ln\rho = 0 \tag{82}$$

$$\frac{\partial \zeta}{\partial t} + \mathrm{div}\left(\frac{1}{m}\rho\nabla\sigma\right) = 0 \tag{83}$$

i.e., they coincide with Eqs. (58) and (59) for a particle in a scalar potential field.

Equations (82) and (83) have the solutions (60) and (61), from which the expressions (62) - (64) follow.

Substituting Eqs. (62) through (64) into equations (76) and (77) for the mean motion of Hamilton-Jacobi ensemble leads to

$$\frac{\partial S}{\partial t} + \frac{1}{2m}\left(\nabla S - \frac{q}{c}\mathbf{A}\right)^2 + U + \frac{\hbar^2}{8m}\frac{|\nabla\rho_0|^2}{\rho_0^2} - \frac{\hbar^2}{4m}\frac{\nabla^2\rho_0}{\rho_0} = 0 \tag{84}$$

$$\frac{\partial \rho_0}{\partial t} + \mathrm{div}\left(\frac{1}{m}\rho_0\nabla S\right) = 0 \tag{85}$$

It is easy to check that equations (84) and (85) are equivalent to one Schrödinger equation for a charged particle in an external electromagnetic field:

$$i\hbar\frac{\partial \psi}{\partial t} = \left[\frac{1}{2m}\left(\frac{\hbar}{i}\nabla - \frac{q}{c}\mathbf{A}\right)^2 + q\varphi\right]\psi \tag{86}$$

where $\psi = \sqrt{\rho}\exp(iS/\hbar)$.

Thus, we have shown that the equations of hydrodynamic type (67) and (40) with a quickly oscillating sign-alternating potential (41) are equivalent to the Schrödinger equation for a spinless particle in an external electromagnetic field: at very high frequencies $\omega$, an average flow of the Hamilton-Jacobi-Schrödinger ensemble is described by the Schrödinger equation (86).

### C. System of interacting particles

Let us consider a system of $N$ interacting particles.

A classical Hamilton-Jacobi ensemble for a system of interacting particles is described by the Euler equation in the configuration space $(\mathbf{r}_1,...,\mathbf{r}_N)$

$$m_i\left(\frac{\partial \mathbf{v}_i}{\partial t}+\sum_{k=1}^{N}(\mathbf{v}_k\nabla_k)\mathbf{v}_i\right)=-\nabla_i U \tag{87}$$

and continuity equation

$$\frac{\partial \rho}{\partial t}+\sum_{k=1}^{N}\nabla_k(\rho\mathbf{v}_k)=0 \tag{88}$$

where $\rho=\rho(\mathbf{r}_1,...,\mathbf{r}_N,t)$ is the density of the Hamilton-Jacobi ensemble in the configuration space $(\mathbf{r}_1,...,\mathbf{r}_N)$, $\mathbf{V}=(\mathbf{v}_1,...,\mathbf{v}_N)$ is the velocity of the point representing the system in the configuration space, $\mathbf{V}(\mathbf{r}_1,...,\mathbf{r}_N,t)$ is the velocity field of the Hamilton-Jacobi ensemble in the configuration space, $U=U(\mathbf{r}_1,...,\mathbf{r}_N,t)$ is the potential energy of the classical interacting particles, depending on the coordinates of all particles, $m_i$ is the mass of the $i$ th particle, and $\nabla_i=\frac{\partial}{\partial \mathbf{r}_i}$, where $i=1,...,N$.

A natural generalization of Eq. (39) to the system of interacting particles given Eq. (87) is as follows:

$$m_i\rho_r\left(\frac{\partial \mathbf{v}_{ri}}{\partial t}+\sum_{k=1}^{N}(\mathbf{v}_{rk}\nabla_k)\mathbf{v}_{ri}\right)=-\rho_r\nabla_i U+\frac{1}{\sqrt{2}}\hbar\omega\cos\omega t\,\nabla_i\rho_r \tag{89}$$

Equation (89) has a solution in the form

$$m_i\mathbf{v}_{ri}=\nabla_i S_r \tag{90}$$

where the function $S_r(\mathbf{r}_1,...,\mathbf{r}_N,t)$ satisfies the Hamilton-Jacobi equation:

$$\frac{\partial S_r}{\partial t}+\sum_{k=1}^{N}\frac{1}{2m_k}|\nabla_k S_r|^2+U-\frac{1}{\sqrt{2}}\hbar\omega\cos\omega t\ln\rho_r \tag{91}$$

As before, we seek the solutions of equations (88), (90) and (91) in the form of (48) and (49). Substituting Eq. (48) into Eqs. (88), (90) and (91) gives

$$\frac{\partial S}{\partial t}+\frac{\partial \sigma}{\partial t}+\sum_{k=1}^{N}\frac{1}{2m_k}|\nabla_k S|^2+\sum_{k=1}^{N}\frac{1}{m_k}(\nabla_k S\nabla_k\sigma)+\sum_{k=1}^{N}\frac{1}{2m_k}|\nabla_k\sigma|^2+U- \\ -\frac{1}{\sqrt{2}}\hbar\omega\cos\omega t\ln\rho-\frac{1}{\sqrt{2}}\hbar\omega(\zeta/\rho)\cos\omega t=0 \tag{92}$$

$$\frac{\partial \rho}{\partial t}+\frac{\partial \zeta}{\partial t}+\sum_{k=1}^{N}\nabla_k\left(\frac{1}{m_k}\rho\nabla_k S+\frac{1}{m_k}\zeta\nabla_k S+\frac{1}{m_k}\rho\nabla_k\sigma+\frac{1}{m_k}\zeta\nabla_k\sigma\right)=0 \tag{93}$$

Averaging equations (92) and (93) over the fast oscillations, while considering Eq. (49) and separating the fast and slow components gives

$$\frac{\partial S}{\partial t} + \sum_{k=1}^{N} \frac{1}{2m_k} |\nabla_k S|^2 + \sum_{k=1}^{N} \frac{1}{2m_k} \langle |\nabla_k \sigma|^2 \rangle + U - \frac{1}{\sqrt{2}\rho} \hbar\omega \langle \zeta \cos\omega t \rangle = 0 \tag{94}$$

$$\frac{\partial \rho}{\partial t} + \sum_{k=1}^{N} \nabla_k \left( \frac{1}{m_k} \rho \nabla_k S + \frac{1}{m_k} \langle \zeta \nabla_k \sigma \rangle \right) = 0 \tag{95}$$

$$\frac{\partial \sigma}{\partial t} + \sum_{k=1}^{N} \frac{1}{m_k} (\nabla_k S \nabla_k \sigma) + \sum_{k=1}^{N} \frac{1}{2m_k} \left( |\nabla_k \sigma|^2 - \langle |\nabla_k \sigma|^2 \rangle \right) - \frac{1}{\sqrt{2}} \hbar\omega \cos\omega t \ln\rho +$$
$$+ \frac{1}{\sqrt{2}\rho} \hbar\omega \left( \langle \zeta \cos\omega t \rangle - \zeta \cos\omega t \right) = 0 \tag{96}$$

$$\frac{\partial \zeta}{\partial t} + \sum_{k=1}^{N} \nabla_k \left( \frac{1}{m_k} \zeta \nabla_k S + \frac{1}{m_k} \rho \nabla_k \sigma + \frac{1}{m_k} \left( \zeta \nabla_k \sigma - \langle \zeta \nabla_k \sigma \rangle \right) \right) = 0 \tag{97}$$

Assuming that the oscillations $\zeta$ and $\sigma$ are small, we are restricted to the linear approximation; using this approximation, Eqs. (96) and (97) can be written in the form

$$\frac{\partial \sigma}{\partial t} - \frac{1}{\sqrt{2}} \hbar\omega \cos\omega t \ln\rho = 0 \tag{98}$$

$$\frac{\partial \zeta}{\partial t} + \sum_{k=1}^{N} \nabla_k \left( \frac{1}{m_k} \rho \nabla_k \sigma \right) = 0 \tag{99}$$

Equations (98) and (99) have the solutions

$$\sigma = \frac{1}{\sqrt{2}} \hbar \sin\omega t \ln\rho \tag{100}$$

$$\zeta = \frac{1}{\sqrt{2}} \frac{\hbar}{\omega} \cos\omega t \sum_{k=1}^{N} \frac{1}{m_k} \nabla_k^2 \rho \tag{101}$$

Using solutions (100) and (101), it is easy to show that

$$\langle \zeta \nabla \sigma \rangle = 0 \tag{102}$$

$$\frac{1}{2m_k} \langle |\nabla_k \sigma|^2 \rangle = \frac{\hbar^2}{8m_k} \frac{|\nabla_k \rho|^2}{\rho^2} \tag{103}$$

$$\frac{1}{\sqrt{2}\rho} \hbar\omega \langle \zeta \cos\omega t \rangle = \frac{\hbar^2}{4} \sum_{k=1}^{N} \frac{1}{m_k} \frac{\nabla_k^2 \rho}{\rho} \tag{104}$$

Substituting Eqs. (102) through (104) into equations (94) and (95) for the averaged parameters of the Hamilton-Jacobi-Schrödinger ensemble leads to

$$\frac{\partial S}{\partial t} + \sum_{k=1}^{N} \frac{1}{2m_k} |\nabla_k S|^2 + U + \sum_{k=1}^{N} \left( \frac{\hbar^2}{8m_k} \frac{|\nabla_k \rho|^2}{\rho^2} - \frac{\hbar^2}{4m_k} \frac{\nabla_k^2 \rho}{\rho} \right) = 0 \tag{105}$$

$$\frac{\partial \rho}{\partial t} + \sum_{k=1}^{N} \nabla_k \left( \frac{1}{m_k} \rho \nabla_k S \right) = 0 \tag{106}$$

It is easy to see that the set of equations (105) and (106) is equivalent to the Schrödinger equation for a system of interacting particles:

$$i\hbar \frac{\partial \psi}{\partial t} = -\sum_{k=1}^{N} \frac{\hbar^2}{2m_k} \nabla_k^2 \psi + U\psi \qquad (107)$$

where $\psi(\mathbf{r}_1,...,\mathbf{r}_N,t) = \sqrt{\rho}\exp(iS/\hbar)$ is the wave function of the system of interacting particles as defined in the configuration space.

Thus, the Hamilton-Jacobi theory (88) and (89) for a system of interacting particles with a quickly oscillating "pressure" (41) contains the quantum mechanics as a limiting case at very high frequencies $\omega$.

## V. DYNAMICS OF INDIVIDUAL QUANTUM PARTICLES

It is well known that the Euler equation for an ideal gas can be obtained from an analysis of the motion and short-range interaction of the multitudes of particles [14]. One can expect that equation (39) can also be obtained from an analysis of the motion of individual quantum particles.

We show below that, within the limits of classical mechanics, one can write the equations of motion of individual particles, which reduce to equations (39) and (40) for the Hamilton-Jacobi ensemble.

Apparently, there are different equations of motion of a classical particle corresponding to equation (39) for the Hamilton-Jacobi ensemble.

Let us consider the formal classical equation of motion of individual particles, leading to equation (39):

$$m\frac{d\mathbf{V}}{dt} = -\nabla U + a\sum_k \nabla \delta(\mathbf{r} - \mathbf{r}_k)$$
$$\frac{d\mathbf{r}}{dt} = \mathbf{V} \qquad (108)$$

where $\mathbf{V}$ is the velocity of an individual particle, $a$ is some function of time, $\mathbf{r}_k$ are the constant vectors, and $\delta(\mathbf{r} - \mathbf{r}_k)$ is the Dirac delta-function, where, for any volume $\Omega$: $\int_\Omega \delta(\mathbf{r} - \mathbf{r}_k)d\mathbf{r} = 1$ if point $\mathbf{r}_k$ lies within the volume $\Omega$ and is otherwise equal to zero.

The vectors $a\nabla\delta(\mathbf{r} - \mathbf{r}_k)$ can be considered potential forces that are created by the $\delta$-sources located at the points $\mathbf{r}_k$ in space.

The first equation (108) is the conventional Newton's second law. Thus, the set of equations (108) describes the motion of a classical particle in an external potential field $U$ in the presence of spatially distributed scattering centers in the form of potential $\delta$-sources.

Let us consider the phase space of the particle $(\mathbf{r}, \mathbf{V})$. The state of the system at any given instant can be represented by a point in the phase space. Let us introduce a statistical ensemble consisting of a set of identical non-interacting systems. The statistical ensemble is represented by the multitude of points in the phase space. We note that the statistical ensemble is defined in the phase space in contrast to the Hamilton-Jacobi ensemble, which is defined in the configuration space.

To describe the statistical ensemble, we use the standard approach of statistical physics [14].

Let us introduce the density of the statistical ensemble in the phase space $P(\mathbf{r}, \mathbf{V}, t)$. If this density is normalized to unity, $P(\mathbf{r}, \mathbf{V}, t)$ can be considered the probability density that a particle has the predetermined position and velocity.

The density $P(\mathbf{r}, \mathbf{V}, t)$ satisfies the continuity equation in phase space, which, for system (108), has the form

$$\frac{\partial P}{\partial t} + (\mathbf{V}\nabla)P + \frac{1}{m}\frac{\partial}{\partial \mathbf{V}}\left(-\nabla U + a\sum_k \nabla \delta(\mathbf{r} - \mathbf{r}_k)\right)P = 0 \tag{109}$$

Here, as usual, the position $\mathbf{r}$ and the velocity $\mathbf{V}$ of the particles are considered to be independent.

Due to the interaction of individual particles with the $\delta$-sources, the distribution function has a small-scale structure with the characteristic spatial scale on the order of the mean distance between adjacent $\delta$-sources. If one assumes that the average distance between adjacent $\delta$-sources is much less than the characteristic spatial scale of motion of the statistical ensemble, one can average equation (109) over a small volume $d\Omega$ with a size much larger than the distance between adjacent $\delta$-sources but much less than the characteristic spatial scales of motion of the statistical ensemble.

By definition of $\delta$-function

$$\int_{d\Omega} P(\mathbf{r}, \mathbf{V}, t)\nabla \delta(\mathbf{r} - \mathbf{r}_k)d\mathbf{r} = -\nabla P(\mathbf{r}_k, \mathbf{V}, t) \tag{110}$$

if the point $\mathbf{r}_k$ lies within the volume $d\Omega$; otherwise, this integral equal to zero.

As a result of the averaging of equation (109) over the volume $d\Omega$, one obtains

$$\frac{\partial \bar{P}}{\partial t} + (\mathbf{V}\nabla)\bar{P} - \frac{1}{m}\left(\nabla U \frac{\partial}{\partial \mathbf{V}}\right)\bar{P} - \frac{1}{m}\frac{\partial}{\partial \mathbf{V}}\left(\frac{a}{d\Omega}\sum_{k \in d\Omega}\nabla P(\mathbf{r}_k, \mathbf{V}, t)\right) = 0 \tag{111}$$

where the summation is over all $\delta$-sources inside the volume $d\Omega$;

$$\overline{P}(\mathbf{r},\mathbf{V},t) = \frac{1}{d\Omega} \int_{d\Omega} P(\mathbf{r},\mathbf{V},t)d\mathbf{r} \tag{112}$$

is the averaged density of statistical ensemble in phase space over the volume $d\Omega$.

If the $\delta$-sources are distributed randomly in space, one can write

$$\frac{1}{d\Omega} \sum_{k \in d\Omega} \nabla P(\mathbf{r}_k,\mathbf{V},t) = n\nabla \overline{P}(\mathbf{r},\mathbf{V},t) \tag{113}$$

where $n = \dfrac{N(d\Omega)}{d\Omega}$ is the density of the $\delta$-sources in physical space and $N(d\Omega)$ is the number of $\delta$-sources inside the volume $d\Omega$.

Then, equation (111) takes the form

$$\frac{\partial \overline{P}}{\partial t} + (\mathbf{V}\nabla)\overline{P} - \frac{1}{m}\left(\nabla U \frac{\partial}{\partial \mathbf{V}}\right)\overline{P} - \frac{1}{m}\frac{\partial}{\partial \mathbf{V}}\left(an\nabla \overline{P}(\mathbf{r},\mathbf{V},t)\right) = 0 \tag{114}$$

The density of the Hamilton-Jacobi ensemble is defined by the expression

$$\rho_r(\mathbf{r},t) = \int \overline{P}(\mathbf{r},\mathbf{V},t)d\mathbf{V} \tag{115}$$

while its velocity $\mathbf{v}_r(\mathbf{r},t)$, which is involved in Eq. (39), is defined by the expression

$$\rho_r \mathbf{v}_r = \int \mathbf{V}\overline{P}(\mathbf{r},\mathbf{V},t)d\mathbf{V} \tag{116}$$

Let us consider a tensor

$$\Pi_{ij} = \int V_i V_j \overline{P}(\mathbf{r},\mathbf{V},t)d\mathbf{V} \tag{117}$$

The velocity of the particles in the statistical ensemble can be decomposed into the mean velocity $\mathbf{v}_r$, which coincides with the flow velocity of the Hamilton-Jacobi ensemble, and the random velocity component $\mathbf{w}$

$$\mathbf{V} = \mathbf{v}_r + \mathbf{w} \tag{118}$$

where

$$\langle \mathbf{w} \rangle = 0 \tag{119}$$

Then, the tensor (117) can be written as

$$\Pi_{ij} = \rho_r \left(v_{ri}v_{rj} + \langle w_i w_j \rangle\right) \tag{120}$$

For isotropic random process $\mathbf{w}$,

$$\langle w_i w_j \rangle = \tau \delta_{ij} \tag{121}$$

where $\tau = \langle w_i^2 \rangle \geq 0$ for any $i = 1,2,3$ is some parameter that is assumed to be constant.

Integrating equation (114) with respect to $\mathbf{V}$ while considering that the probability of infinite velocities is equal to zero gives

$$\frac{\partial \rho_r}{\partial t} + \operatorname{div}(\rho_r \mathbf{v}_r) = 0 \tag{122}$$

i.e., the continuity equation (40).

Multiplying equation (114) by **V** and integrating it with respect to **V** while considering Eqs. (115) through (122) gives

$$m\rho_r\left(\frac{\partial \mathbf{v}_r}{\partial t} + (\mathbf{v}_r \nabla)\mathbf{v}_r\right) = -\rho_r \nabla U - (an+\tau)\nabla \rho_r \qquad (123)$$

By comparing equation (123) with equation (39), one concludes that they coincide if we take

$$an + \tau = -\frac{1}{\sqrt{2}}\hbar\omega\cos\omega t \qquad (124)$$

Thus, we have shown, at least formally, that, if the motion of a particle is described using the classical dynamical equation (108) with the parameters (124), the Hamilton-Jacobi ensemble of the particles is described by the equations of hydrodynamic type (39) and (40). In doing so, the mean motion of the Hamilton-Jacobi ensemble is described by the Schrödinger equation.

## VI. CONCLUDING REMARKS

We have shown that the hydrodynamic equations (39) and (40) with a quickly oscillating sign-alternating pressure (41) are equivalent to the Schrödinger equation in the sense that at very high frequencies $\omega$, the flow of the Hamilton-Jacobi-Schrödinger ensemble, defined by the equations (39) and (40), after averaging over the fast oscillations, is described by the Schrödinger equation. It follows that the Schrödinger equation in this model is an approximate equation and describes only the mean motion of Hamilton-Jacobi-Schrodinger ensemble, while equations (39) and (40) describe the motion of the ensemble in detail, which contains fast oscillations. In particular, Eqs. (39) and (40) may contain a number of fine physical effects that are absent in the Schrödinger equation because they are lost due to averaging over the fast oscillations. It is interesting to investigate these effects.

We have shown that the Schrödinger equation corresponds to the limit of very high frequencies, $\omega$. What is this frequency? One can assume that this frequency is "a natural frequency" of the particle associated with its rest mass: $\omega = mc^2/\hbar$. At least for the main non-relativistic quantum particles (i.e., electrons, protons, neutrons), this frequency can be considered very high in comparison to the frequencies of all processes involving these particles.

The Schrödinger equation, in the end, can be obtained from the equations of motion of classical particles (108), which is of great interest from the perspective of the problem of interpreting quantum mechanics [5,6].

Let us consider a dynamic interpretation of Newton's equation (108).

Let there be a multitude of $\delta$-sources distributed in space with a concentration $n$ that create a potential field $a\sum_{k}\delta(\mathbf{r}-\mathbf{r}_k)$; the amplitude of the sources is defined by Eq. (124) and oscillates with a high frequency $\omega$. In addition, there is a classical potential field $U(\mathbf{r},t)$ in space. If a classical particle moves in this space, its motion is described by the equations of classical mechanics (108), while the Hamilton-Jacobi ensemble corresponding to this system is described by equations (39) and (40). The ensemble executes a complex motion that consists of the mean (observed) motion and the fast oscillations with frequency $\omega$. As shown above, the mean (observed) parameters of such an ensemble are described by the Schrödinger equation.

This model resembles the famous pinball game in which the ball moves randomly between small repulsive obstacles. For this reason, this model can be called "a pinball model". Obviously, the pinball model resonates with the stochastic model of quantum motion [15,16]. We are not going to discuss the nature of these $\delta$-sources; we note only that they can be related to the physical vacuum.

## References

[1] L.D. Landau and E.M. Lifshitz, *Mechanics* (Butterworth-Heinemann, 3rd ed., 1976), Vol. 1.

[2] H. Goldstein, C. P. Poole, and J. L. Safko, *Classical Mechanics* (Addison Wesley, 3rd edition, 2001).

[3] M. Born, *Z. Phys.* 37, 863 (1926). (Reprinted and translated in *Quantum Theory and Measurement*, edited by J. A. Wheeler and W. H. Zurek, (Princeton University Press, Princeton, NJ, 1963).

[4] A. Messiah, *Quantum Mechanics* (Dover Publications Inc. New York, 1999).

[5] R. Omnès, *The Interpretation of Quantum Mechanics* (Princeton University Press, Princeton, N.J. 1994).

[6] G. Auletta, *Foundations and interpretation of quantum mechanics* (World Scientific Publishing Co. Pte. Ltd, 2001).

[7] D. Bohm, Phys. Rev. 85: 166–179 (1952).

[8] D. Bohm, Phys. Rev. 85: 180–193 (1952).

[9] D. Dürr and S. Teufel, *Bohmian Mechanics. The Physics and Mathematics of Quantum Theory* (Springer-Verlag Berlin Heidelberg 2009).

[10] L.D. Landau, E.M. Lifshitz, *The Classical Theory of Fields* (Butterworth-Heinemann, 4th ed., 1975), Vol. 2.

[11] L.D. Landau and E.M. Lifshitz, *Quantum Mechanics: Non-Relativistic Theory* (Pergamon Press, 3rd ed., 1977), Vol. 3.

[12] L.D. Landau and E.M. Lifshitz, *Fluid Mechanics* (Butterworth-Heinemann, 2nd ed., 1987), Vol. 6

[13] V. B. Krapchev, *Phys. Rev. Lett.* 42, 497 (1979)

[14] L.P. Pitaevskii and E.M. Lifshitz, *Physical Kinetics* (Pergamon Press, 1st ed., 1981), Vol. 10.

[15] E. Nelson, *Phys. Rev.* 150: 1079–1085 (1966).

[16] E. Nelson, *Quantum Fluctuations* (Princeton Series in Physics, Princeton University Press, Princeton, NJ, 1985).